\documentclass[twocolumn]{aastex61}
\pdfoutput=1 
\usepackage{amsmath,amstext}
\usepackage[T1]{fontenc}
\usepackage{apjfonts} 
\usepackage[figure,figure*]{hypcap}

\shorttitle{The Hunt for IMBHs in the JWST Era}
\shortauthors{J.M. Cann et al.}

\begin{document}

\title{The Hunt for Intermediate Mass Black Holes in the JWST Era}

\author{Jenna M. Cann}
\affiliation{George Mason University, Department of Physics and Astronomy, MS3F3, 4400 University Drive, Fairfax, VA 22030, USA}
\affiliation{National Science Foundation, Graduate Research Fellow}
\author{Shobita Satyapal}
\affiliation{George Mason University, Department of Physics and Astronomy, MS3F3, 4400 University Drive, Fairfax, VA 22030, USA}
\author{Nicholas P. Abel}
\affiliation{MCGP Department, University of Cincinnati, Clermont College, Batavia, OH 45103, USA}
\author{Claudio Ricci}
\affiliation{N\'ucleo de Astronom\'ia de la Facultad de Ingenier\'ia, Universidad Diego Portales, Av. Ej\'ercito Libertador 441, Santiago, Chile}
\affiliation{Instituto de Astrofisica, Facultad de Fisica, Pontificia Universidad Catolica de Chile, Casilla 306, Santiago 22, Chile}
\affiliation{Chinese Academy of Sciences South America Center for Astronomy and China-Chile Joint Center for Astronomy, Camino El Observatorio 1515, Las Condes, Santiago, Chile}
\affiliation{Kavli Institute for Astronomy and Astrophysics, Peking University, Beijing 100871, China}
\author{Nathan J. Secrest}
\affiliation{U.S. Naval Observatory, 3450 Massachusetts Avenue NW, Washington, DC 20392, USA}
\author{Laura Blecha}
\affiliation{University of Florida, Department of Physics, P.O. Box 118440, Gainesville, FL 32611-8440}
\author{Mario Gliozzi}
\affiliation{George Mason University, Department of Physics and Astronomy, MS3F3, 4400 University Drive, Fairfax, VA 22030, USA}

\begin{abstract}
Intermediate mass black holes (IMBHs), with masses between $100 - 10^5$ M$_{\odot}$, represent the link between stellar mass black holes and the supermassive black holes that reside in galaxy centers.  While these IMBHs are crucial to our understanding of black hole seed formation, black holes of less than $\approx 10^4$ M$_{\odot}$ eluded detection by traditional searches.  Observations of the infrared coronal lines (CLs) offer us one of the most promising tools to discover IMBHs in galaxies. We have modeled the infrared emission line spectrum that is produced by gas photoionized by an AGN radiation field and explored for the first time the dependence of the infrared CL spectrum on black hole mass over the range of $10^2$ M$_\odot - 10^8$ M$_{\odot}$. We show that infrared CLs are expected to be prominent in the spectra of accreting IMBHs and can potentially be a powerful probe of the black hole mass in AGNs. In particular, emission line ratios involving ions with the highest ionization potentials with respect to those with lower ionization potentials, such as [SiXI]/[SiVI], [SiIX]/[SiVI], and [FeXIII]/[FeVI], vary by as much as seven orders of magnitude over the mass range explored in our calculations, with the highest ratios corresponding to the lowest mass black holes. We identify key emission line ratios that are most sensitive to black hole mass spanning six orders of magnitude in black hole mass. While variations in accretion rate and the physical parameters of the gas can also affect the CL spectrum, we demonstrate that the effect of black hole mass is likely to be the most dramatic over the mass range explored in our models. With the unprecedented sensitivity of JWST, a large number of CLs will be detected for the first time in numerous galaxies, providing important insight into the existence and properties of IMBHs in the local universe, potentially revolutionizing our understanding of this class of object. 

\end{abstract}


\section{Introduction}

Black holes with masses from a million to up to a billion times the mass of the sun are now known to be ubiquitous in the centers of galaxies, and their masses appear to correlate with properties of the host in which they reside \citep[e.g.,][]{magorrian1998, gebhardt2000, gultekin2009, mcconnell2013}.  There is now a growing number of black holes found with masses between $10^5 - 10^6$ M$_{\odot}$ \citep{greene2004,greene2007, greene2010, jiang2011, xiao2011, dong2012, reines2013, baldassare2015}. However, very little is known about the existence, properties, host galaxy demographics, and scaling relations of black holes with masses less than $\approx10^5$M$_{\odot}$.  In fact, there is currently no \textit{direct} evidence for black holes with masses between $\approx 60$ M$_{\odot}$ and $\approx10^4$M$_{\odot}$.  Black holes in this "mass desert" in the local universe are of significant astrophysical interest.  This is because they are potential analogs of the original seed black holes formed at redshifts z $>$ 15.  The mass function and occupation fraction of these "intermediate mass black holes (IMBHs)" in the local universe hold vital clues into the origins of supermassive black holes (SMBHs), potentially allowing us to discriminate between lower mass seeds formed from stellar remnants or massive seeds formed directly out of the collapse of dense gas \citep{volonteri2009, volonteri2010, vanwassenhove2010, greene2012, reines2016, mezcua2017}.  Moreover, mergers between black holes in this mass range are one of the most promising sources of gravitational waves (GWs) detectable with the Laser Interferometer Space Antenna \citep[LISA;][]{amaroseoane2012,amaroseoane2013}, yet black hole pairs in this mass range in the local universe have not yet been identified and their merger rate is unknown.  Apart from their importance in understanding the origin of SMBHs, IMBHs are of intrinsic interest in studying the physics of accretion in the low-mass regime, where some of the fundamental signposts of accretion activity, such as the broad line region \citep{elitzur2009, chakravorty2014} and the torus \citep{krolik1988} may disappear.  Finding a population of IMBHs, measuring their masses, determining their merger rates, and understanding their connection to their host galaxies is therefore of fundamental astrophysical importance. \par

Unfortunately, finding and studying the properties of IMBHs is challenging since the sphere of influence of even a $10^5$M$_{\odot}$ black hole at a distance of 10 Mpc is only approximately $0.01$".  Detecting a population of IMBHs through resolved kinematics is therefore currently observationally impossible.  A significant sample of IMBHs can therefore only be detected if they are accreting.  However, accretion activity for such low mass black holes is also challenging to detect.  This is because accreting IMBHs are likely to be found in the centers of low-mass galaxies, where star formation in the host galaxy can dominate the optical spectrum \citep{trump2015} and gas and dust can obscure the central engine at optical, and even X-ray, wavelengths.  Indeed, recent {\it NuSTAR} observations are revealing a growing number of nearby low luminosity active galactic nuclei (AGNs) that are Compton thick \citep[e.g.,][]{annuar2015, annuar2017, ricci2016}.  Moreover, even if highly accreting and unobscured, AGNs powered by IMBHs will have low luminosities. For example, a  $10^3$M$_{\odot}$ black hole radiating at its Eddington rate will have a bolometric luminosity of only $10^{41}$~erg~s$^{-1}$. The X-ray luminosities of these low luminosity AGNs will be low and can be comparable to, and therefore indistinguishable from, a population of high mass X-ray binaries in the host galaxy.  This problem is exacerbated in low mass galaxies which have lower metallicities, where X-ray emission from high mass X-ray binaries is enhanced \citep[e.g.,][]{mapelli2009, fragos2013}.  Likewise, the radio emission from a potential AGN in a low mass galaxy can be comparable to, and indistinguishable from, a compact nuclear starburst \citep{condon1991}, making it impossible to uniquely identify accretion around IMBHs with radio observations alone.  While mid-infrared color-selection is a powerful tool in uncovering obscured AGNs \citep[e.g.,][Satyapal et al. in press]{lacy2004, stern2005, donley2012, stern2012, assef2013, mateos2012}, it is well-known that this method fails in galaxies where the luminosity of the stellar emission from the host galaxy is comparable to, or exceeds, that from the AGN.  Young starbursts can also mimic the mid-infrared colors of luminous AGNs \citep{hainline2016}, although only under extreme conditions (Satyapal et al. 2018 in press). Even broad emission lines in the optical spectrum, usually a hallmark signature of an AGN powered by a massive black hole, are associated with supernova activity for the majority of cases in dwarf galaxies \citep{baldassare2016}, further emphasizing the limitations of optical studies in the hunt for IMBHs in the local universe, although these sources have narrow emission lines consistent with star forming galaxies.  Finally, even the standard optical emission lines used to classify galaxies dominated by AGNs using the Baldwin-Phillips-Terlevich (BPT) diagram \citep{baldwin1981} has been established only for higher mass black holes and are also affected by the presence of shocks \citep[e.g.,][]{kewley2001,kauffmann2003}.  It is not yet known whether these diagnostic diagrams are robust AGN indicators in the IMBH mass range. We explore these diagnostics in a future paper (Cann et al. 2018 in prep). \par

Given the challenges associated with finding IMBHs, current searches have yielded relatively small numbers of candidates. There are several candidate IMBHs reported in globular clusters, where some firm upper limits on the black hole masses have been obtained \citep[e.g.,][]{maccarone2005,maccarone2007,strader2012,lutzgendorf2012,lutzgendorf2013,wrobel2016,kiziltan2017,perera2017MNRAS}. Searches for AGNs based on optical and X-ray studies in low mass galaxies have yielded a small but slowly growing sample of candidate active IMBHs \citep[e.g.,][]{filippenko2003, barth2004, satyapal2007, greene2007, satyapal2008, satyapal2009, dewangan2008, shields2008, ghosh2008, izotov2008, desroches2009, gliozzi2009,mcalpine2011, jiang2011, reines2011, secrest2012, ho2012, dong2012, araya2012, secrest2013, simmons2013, reines2013, coelho2013, schramm2013, bizzocchi2014, reines2014, maksym2014, moran2014, yuan2014a,secrest2015,miller2015,mezcua2016,lemons2015, baldassare2015,pardo2016,baldassare2017,chen2017}. However, the fraction of low mass galaxies with evidence of accretion identified through these techniques is extremely small and the mass range probed is still above $\approx 10^4$ M$_{\odot}$.


Observations of the infrared fine structure lines offer us one of the only definitive tools to discover buried AGNs in dusty galaxies.  As has been shown in previous works, AGNs show prominent high-excitation fine structure line emission, whereas starburst and normal galaxies are characterized by a lower excitation spectra characteristic of HII regions ionized by young stars \citep[e.g.,][]{genzel1998, sturm2002,satyapal2004}.  Fine structure lines from ions with ionization potentials greater than $\approx 70$ eV \citep[Satyapal et al. in prep]{abel2008}, the so-called "coronal" lines (CLs), cannot be easily produced in HII regions surrounding young stars, the dominant energy source in starburst galaxies, since even hot massive stars emit very few photons with energy sufficient for the production of these ions.  The power of these diagnostics in finding buried AGN has been strikingly demonstrated by the {\it Spitzer} mission through the discovery of a population of AGNs in galaxies that display optically "normal" nuclear spectra \citep[e.g.,][]{lutz1999,satyapal2007,satyapal2008,satyapal2009,goulding2009}.  Given that AGNs powered by IMBHs will be low luminosity, reside in low mass galaxies with potentially enhanced star formation \citep{trump2015}, and may have enhanced obscuration \citep{geha06} compared with AGNs in higher mass galaxies, an IR {\it spectroscopic} study is crucial to provide unambiguous proof of an AGN and study its properties in this class of objects. \par

With the advent of the {\it James Webb Space Telescope (JWST)}, infrared spectroscopic observations with unprecedented sensitivity will be possible.  These observations can potentially uncover IMBHs in large samples of galaxies, possibly revolutionizing our understanding of this class of objects.  Not only can infrared CLs uniquely identify accretion activity from low-luminosity AGNs residing in dusty star forming hosts, but since the ionization potentials of the associated ions are in the ultraviolet to soft X-ray regime, at wavelengths that are not directly observable due to Galactic absorption, the infrared CLs can potentially be used to indirectly reconstruct the shape of the spectral energy distribution (SED) of the AGN at wavelengths where the emission from the accretion disk is expected to peak \citep[e.g.,][]{lyndenbell1969,shakura1973,netzer1985}, potentially allowing us to gain insight into the accretion properties of the black hole and its mass.  As the black hole mass decreases, the Schwarzchild radius decreases, and in response, the temperature of the surrounding accretion disk increases.  The shape of the ionizing radiation field therefore changes with black hole mass, which in turn will impact the emission line spectrum at both optical and infrared wavelengths. Not only do infrared fine structure lines carry the advantage of being less sensitive to dust extinction, the excitation energies corresponding to the transitions are small compared to the ambient nebular temperature, and so the infrared line ratios of two different stages of ionization produced by the same element are only weakly dependent on the electron temperature of the gas in which they are produced. They are therefore the ideal tool to study the nature of the ionizing radiation field in the dust enshrouded nuclear regions of galaxies. \par

In this paper, we explore the diagnostic potential of infrared CLs in the study of AGNs, investigating their behavior as a function of black hole mass. In Section 2, we describe our model for the AGN continuum and discuss our photoionization calculations.  In Section 3, we show the predicted emission line strengths.  We summarize our main findings in Section 4.

\section{Theoretical Calculations}

Coronal line emission can in principle originate in gas photoionized by a hard ionizing continuum \citep[e.g.,][]{shields1975,grandi1978,korista1989,ferguson1997} or a hot ( $\approx 10^{6}$ K) collisionally ionized plasma \citep[e.g.,][]{oke1968,nussbaumer1970}. Several studies suggest that the main driver of the CL emission is photoionization by a hard radiation field \citep[e.g.,][]{oliva1994,whittle2005,schlesinger2009,mazzalay2010} with possibly some contribution from shocks \citep[e.g.,][]{rodriguez2006, geballe2009}. In this work, we assume that the CL emission is produced in gas ionized by the AGN radiation field and explore the dependence on the black hole mass. In order to model the infrared emission line spectrum of gas ionized by an AGN, we used version c17 of the spectral synthesis code \textsc{Cloudy} \citep{ferland2017}. The emission line spectrum depends on several factors, including the shape of the ionizing radiation field, the geometrical distribution of the gas, the chemical composition and grain properties, the gas properties, and the equation of state. In this work, we specifically explore the behavior of the CL emission. We therefore assume that the ionizing radiation field incident on the gas is produced exclusively by an AGN continuum since the ionizing radiation from a stellar population does not contribute to the ionization of these lines. This is demonstrated in a separate publication (Satyapal 2018 in prep). We note that our calculations are not intended to model the CL ratios of any one particular AGN, but to explore the dependence of the CL emission on black hole mass under the assumption that the CL emission is produced by gas photoionized by a simple mass-dependent generic AGN continuum. Below we discuss the details of our model parameters.

\subsection{The AGN Continuum}
We assume that the AGN continuum consists of three components: an accretion disk, Comptonized X-ray radiation in the form of a power law, and an additional component seen in the X-ray spectrum of most AGNs, often referred to as the "soft excess component". 

\subsubsection{The Accretion Disk}
Multiwavelength observations of quasars suggest that the continuum emission from AGNs peaks in the ultraviolet part of the electromagnetic spectrum \citep[e.g.,][]{shields1978,elvis1986,laor1990}. This emission, often referred to as the "Big Blue Bump"(BBB) is attributed to the emission from the accretion disk around the black hole. In our models, we assume that the accretion disk is a simple geometrically thin, optically thick disk \citep{shakura1973}, where the emission from the disk is approximated by the superposition of blackbodies at temperatures corresponding to the  different disk annuli at radius R, with the temperature as a function of radius R given by \citet{peterson07, frank2002}:

\begin{equation} \label{Eq 1}
T = 6.3 \times 10^{5} \bigg( \frac{\dot m}{\dot m_{Edd}} \bigg)^{1/4} \bigg( \frac{M_{BH}}{10^{8}M_{\odot}} \bigg)^{-1/4}\bigg( \frac{R}{R_s} \bigg)^{-3/4} K
\end{equation}

where $\dot m_{Edd}$ and $\dot m$ are the Eddington and actual accretion rates, respectively, M$_{BH}$ is the mass of the black hole, and R$_s$ is the Schwarzchild radius. \par
\begin{figure}
\includegraphics[width=0.5\textwidth]{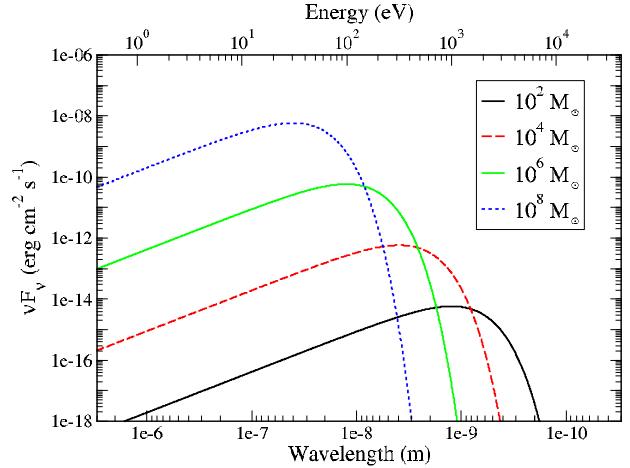}
\caption{Accretion disk SEDs from black holes with masses that span 10$^{2}-10^{8}$ M$_{\odot}$ accreting at 0.1 $\dot m_{Edd}$. The units of the vertical axis are set for illustrative purposes by assuming a distance of $30$ Mpc. We have labeled the location of the ionization potentials of several coronal lines accessible by {\it JWST} by the dotted vertical lines. As can be seen, the shift in disk temperature for the range of black hole masses explored in this work significantly moves the location of the "Big Blue Bump" relative to the ionization energies of key coronal lines observable by {\it JWST}.}
\label{figure_label1}
\end{figure}

In most observed SMBHs, where masses range from 10$^{6}$-10$^{9}$ M$_{\odot}$, this temperature change is not significant.  Indeed, the inner temperature of the accretion disk changes by only a factor of 2 when the black hole mass changes by three orders of magnitude. Because of this, most studies do not take temperature-dependent changes in the accretion disk SED shape into account.  When looking at wider range of black hole masses, however, this temperature difference can cause the peak of the accretion disk spectrum to move dramatically (Figure \ref{figure_label1}). Moreover, because the ionization potentials of the coronal lines fall at energies very close to the peak of the BBB, changes in black hole mass will strongly affect the coronal line emission spectrum emitted by the surrounding gas, potentially providing a diagnostic of the black hole mass.  

In this study, the accretion disk spectrum was modeled using the \textsc{diskbb} model \citep{mitsuda1984, makishima1986} of XSpec v12.9.0\footnote{\url{http://www.heasarc.gsfc.nasa.gov/docs/xanadu/xpec}}  \citep{arnaud96}. To set the temperature at the inner disk radius, we used Equation \ref{Eq 1} and varied the black hole mass from 10$^{2}-10^{8}$ M$_{\odot}$ and assumed that the black hole is accreting at 0.1 $\dot m_{Edd}$. The accretion disk SED from this model is shown in Figure \ref{figure_label1}. Over the range of black hole mass explored in this work the peak of the BBB varies from $13$ eV to $424$ eV. 

\subsubsection{The Power Law Component}

Observations of AGN spectra show a high energy component that can be well-approximated by a power law, believed to be caused by Comptonization of seed photons produced by the disk \citep{svensson1999} by energetic electrons in a hot corona \citep{krolik1999}.  In this study, we adopt a spectral index of 0.8 \citep[e.g.,][]{wilkes1987, grupe2006} and an $\alpha_{OX}$ ratio of 1.2 \citep{netzer1993} for this power law component.  We assume a high energy cutoff of 100 keV.  As the emission lines we are analyzing have ionization potentials of less than or equal to 500 eV, the detailed shape of the AGN SED at the highest energies will not play a significant role in our calculations. 

\subsubsection{The Soft Excess}

An additional component to the AGN has been observed as excess radiation between 100 to 300 eV that cannot be explained by simply extending the power law to lower energies.  This emission, referred to as the "soft excess", has been detected in over 50\% of Seyfert 1 galaxies \citep{halpern1984, turner1989}, though recent studies of local Seyfert 1s suggest that the fraction can reach up to 70-90\% \citep[e.g.,][]{piconcelli05, bianchi09, scott2012, boissay16}.  The origin of this emission is still a topic of debate, with three primary schools of thought being: a Comptonized disk \citep{czerny87, done2012, jin2017}, blurred ionized reflection \citep{fabian2005, crummy06}, or relativistic smeared absorption \citep{gierlinski04, middleton07}.

\par
While  the origin of the soft excess is not clear, this spectral component is remarkably well fit with a blackbody with a roughly constant temperature of $\approx$ $100-200$ eV \citep[e.g.,][]{gierlinski04, ricci2017}. In this work, we model the soft excess phenomenologically as a blackbody with a temperature of 150 eV.  We assume that the ratio of the luminosity of the soft-excess component to the luminosity of the power law component is constant with respect to black hole mass and adopt a ratio consistent with what is typically observed (1) \citep{vignali04, piconcelli05, chakravorty2012}.  Little variation in the soft excess emission with respect to mass is seen \citep{gierlinski04, boissay16}, but only a small range of black hole masses have been explored. In Figure 2, we plot the incident continuum, including all three components, for our $10^7$M$_{\odot}$ model. For black holes with mass $<10^3$M$_{\odot}$, the emission from the disk shifts into the soft X-rays and dominates over the soft excess.  Note that because the ionization potentials of the coronal lines extend into the soft X-rays, it is important to include this component into our AGN continuum. We emphasize that we have not attempted to present a model finely tuned to match the properties of any one particular AGN.  The strength and shape of the continuum in the $100-400$ eV range is important in replicating observed CL line ratios. The generic AGN model adopted in this work predicts the relative behavior of the CL ratios as a function of BH mass, and demonstrates which line ratios are the most sensitive to BH mass. 


\begin{figure}
\includegraphics[width=0.5\textwidth]{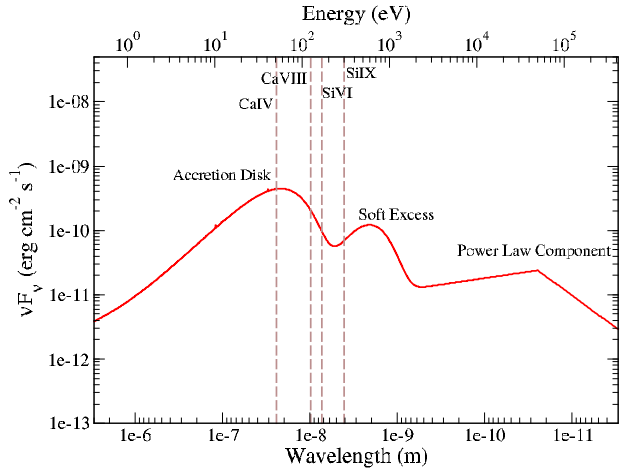}
\caption{SED from black hole of mass 10$^7$ M$_{\odot}$ with soft excess and power law components. The units of the vertical axis are set by assuming a distance of 30Mpc. We have labeled the location of the ionization potentials of several coronal lines accessible by {\it JWST} by the dotted vertical line. As can be seen, several coronal lines have an ionization potential that could be greatly affected by the presence of a soft excess component.}
\label{figure_label2}
\end{figure}

\subsection{Geometrical Distribution of the Gas}

We assume a one-dimensional spherical model with a closed geometry, where the cloud is between the observer and the continuum source, and the ionization parameter and gas density are allowed to vary.  The ionization parameter, $U$, is defined as the dimensionless ratio of the incident ionizing photon density to the hydrogen density:
\begin{equation}
U=\frac{\phi_H}{n_H}=\frac{Q(H)}{4\pi R^{2}n_{H}c}
\end{equation}

 where $\phi_H$ is the flux of hydrogen-ionizing photons striking
the illuminated face of the cloud per second, $n_{H}$ is the hydrogen density, $c$ is the speed
of light, and $Q(H)$ is the number of hydrogen ionizing photons
striking the illuminated face per second. In our calculations, U is allowed to vary from  $10^{-4}$ to $10^{-1}$.   On average, U is $\sim10^{-3}$ based on observations of the optical emission lines in local galaxies and HII regions \citep{dopita2000, moustakas2010}, but can be higher in ULIRGS \citep{abel09} or high-redshift galaxies \citep{pettini01, brinchmann08, maiolino08, hainline09, erb2010}, and in the CL region. We therefore explored a large dynamic range of ionization parameters in this work.
\par

\subsection{Physical State of the Gas and Dust}

We chose the gas and dust abundances in our calculations to be consistent with the local interstellar medium (ISM). The effects of variations in metallicity on the computed spectrum will be explored in future work. We assume constant density, and vary the hydrogen density of the gas from $\log(n_H/cm^{3})$ = $1.5-3.5$, in units of 1.0 dex. We include thermal, radiation, and turbulent pressure in our models with the turbulent velocity set to 5.0 km s$^{-1}$. Our calculations extend only to the Hydrogen ionization front since all of the lines are produced in the ionized region of the cloud.
\par

Given the full set of black hole masses, ionization parameters, and gas densities, we computed a total of 507 simulations. For each model, we computed the emergent spectrum of all emission lines with ionization potentials in excess of 70~eV that are within $1-30$ $\mu$m.  We explored the behavior of only the most prominent lines that can potentially be detected by {\it JWST}, the brightest of which are listed in Table \ref{tab1}.

\begin{table*}[t]
\caption{Brightest Coronal Lines in the $1-30~\mu$m Range}

\centering
\begin{tabular}{lcccc}
\hline
\hline
\noalign{\smallskip}
Line & Transition & Wavelength & Ionization Potential & Critical Density\\
 & &   ($\mu$m) & (eV) & (cm$^{-3}$)\\
\noalign{\smallskip}
\hline
\noalign{\smallskip}

 [Fe VI] &  $^4$P$_{1/2} - ^2$D2$_{3/2}$ & 1.01089  & 75 & $1.070 \times 10^7$\\
 \hline
 [Fe XIII]  & $^3$P$_0 - ^3$P$_1$ & 1.07462 & 331 & $6.498 \times 10^8$ \\
 \hline
 [Si X] &  $^2$P$_{1/2} - ^2$P$_{3/2}$ & 1.43008  & 351 & $1.300 \times 10^8$ \\
 \hline
 [Si XI] &  $^3$P$_1 - ^3$P$_2$ & 1.93446 & 523 & $1.129 \times 10^8$ \\
 \hline
 [Si VI] & $^2$P$_{1/2} - ^2$P$_{3/2}$ & 1.96247 & 167 & $3.022 \times 10^8$ \\ 
 \hline
 [Al IX] & $^2$P$_{1/2} - ^2$P$_{3/2}$ & 2.04444  & 285 & $9.806 \times 10^7$ \\
 \hline
 [Ca VIII] & $^2$P$_{1/2} - ^2$P$_{3/2}$ & 2.32117 & 127 & $5.012 \times 10^6$ \\
 \hline
 [Si VII] & $^3$P$_{2} - ^3$P$_{1}$ & 2.48071 & 205 & $1.101 \times 10^8$ \\
 \hline
 [Al V] & $^2$P$_{1/2} - ^2$P$_{3/2}$ & 2.90450  & 120 & $1.067 \times 10^8$ \\
 \hline
 [Ca IV] & $^2$P$_{1/2} - ^2$P$_{3/2}$ & 3.20610 & 51 & $1.259 \times 10^7$ \\
 \hline
 [Al VI] & $^3$P$_1 - ^3$P$_2$ & 3.65932 &  154 & $3.400 \times 10^6$ \\
 \hline
 [Si IX] & $^3$P$_1 - ^3$P$_0$ & 3.92820 & 304 & $1.114 \times 10^7$ \\
 \hline 
 [Mg IV] & $^2$P$_{1/2} - ^2$P$_{3/2}$ & 4.48712 & 80 & $1.285 \times 10^7$ \\
 \hline
 [Ar VI] &$^2$P$_{1/2} - ^2$P$_{3/2}$ & 4.52800 & 75 & $7.621 \times 10^5$ \\
 \hline
 [Mg VII] & $^3$P$_{1} - ^3$P$_{2}$ & 5.50177 & 187 & $4.580 \times 10^6$ \\
 \hline
 [Mg V] & $^3$P$_{2} - ^3$P$_{1}$  & 5.60700 & 109 & $5.741 \times 10^6$ \\
 \hline
 [Si VII] & $^3$P$_{1} - ^3$P$_{0}$ & 6.51288 & 205 & $1.254 \times 10^7$ \\ 
 \hline
 [Ne VI] & $^2$P$_{1/2} - ^2$P$_{3/2}$ & 7.64318 & 126 & $3.573 \times 10^5$ \\
 \hline
 [Fe VII] & $^3$F$_{3} - ^3$F$_{4}$ & 7.81037 & 99 & $1.733 \times 10^6$ \\
 \hline
 [Ar V] & $^3$P$_{1} - ^3$P$_{2}$ & 7.89971 & 60 & $2.025 \times 10^5$ \\
 \hline
 [Mg VII] & $^3$P$_1 - ^3$P$_0$ & 9.00655 & 187 & $2.742 \times 10^6$ \\
 \hline
 [Na IV] & $^3$P$_{2} - ^3$P$_{1}$ & 9.03098 & 72 & $1.313 \times 10^6$ \\
 \hline
 [Fe VII] & $^3$F$_{2} - ^3$F$_{3}$ & 9.50763 & 99 & $1.858 \times 10^8$ \\
 \hline
 [Fe VI] & $^4$F$_{7/2} - ^4$F$_{9/2}$ & 12.3074 & 75 & $3.020 \times 10^5$ \\
 \hline
 [Ar V] & $^3$P$_{0} - ^3$P$_{1}$ & 13.0985 & 60 & $9.516 \times 10^4$ \\
 \hline
 [Ne V] & $^3$P$_{1} - ^3$P$_{2}$ &  14.3228 & 97 & $4.792 \times 10^4$ \\
 \hline
 [Fe VI] & $^4$F$_{5/2} - ^4$F$_{7/2}$ & 14.7670 & 75 & $2.996 \times 10^5$\\
 \hline
 [Fe VI] & $^4$F$_{3/2} - ^4$F$_{5/2}$ & 19.5527 & 75 & $1.456 \times 10^5$ \\
 \hline
 [Ne V] & $^3$P$_{0} - ^3$P$_{1}$ & 24.2065 & 97 & $2.656 \times 10^4$ \\

\noalign{\smallskip}
\hline
\noalign{\smallskip}

\end{tabular}
\tablecomments{Column 4: Ionization potential corresponding to each coronal line.}
\label{tab1}
\end{table*}

\section{Results}

\begin{figure}
\includegraphics[width=0.45\textwidth]{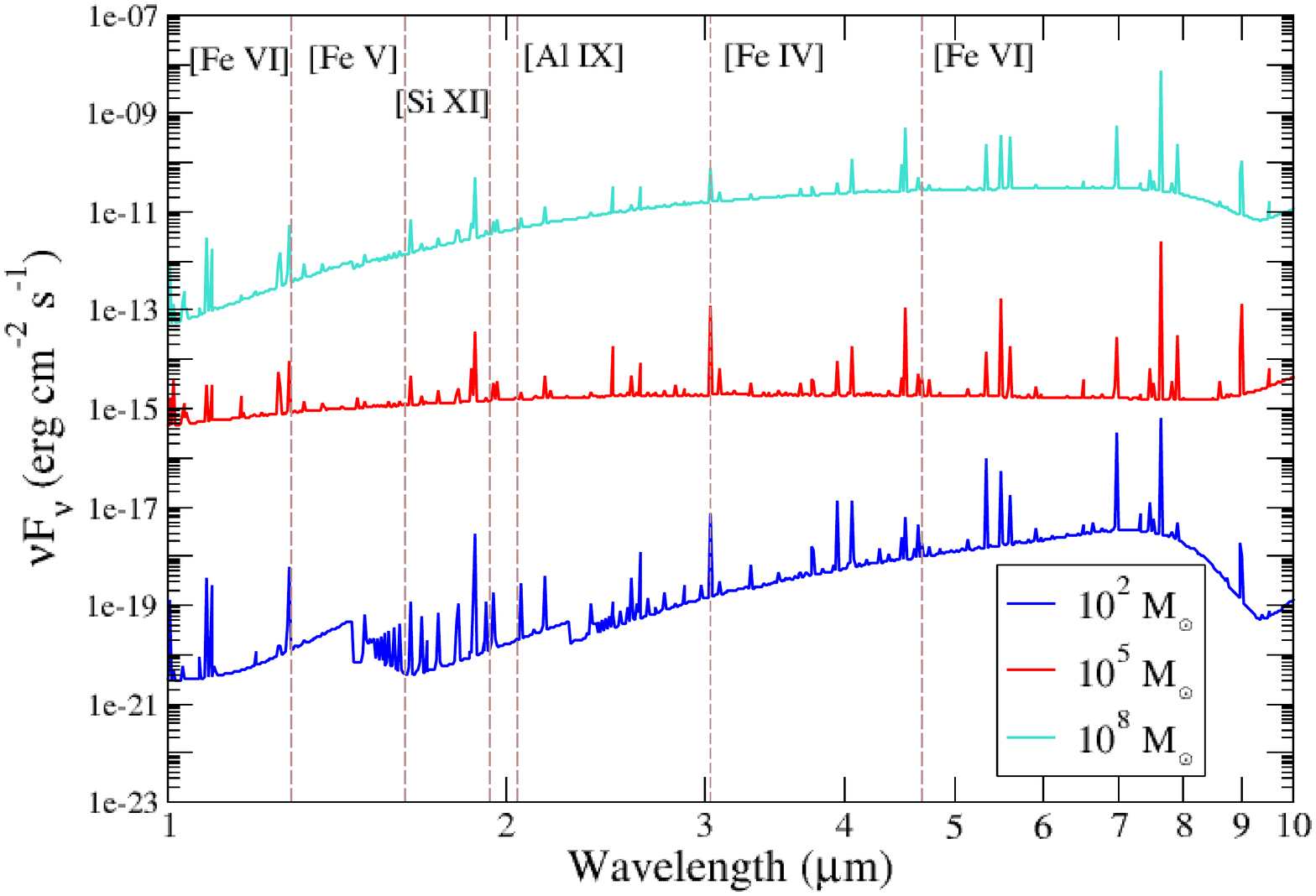}
\includegraphics[width=0.45\textwidth]{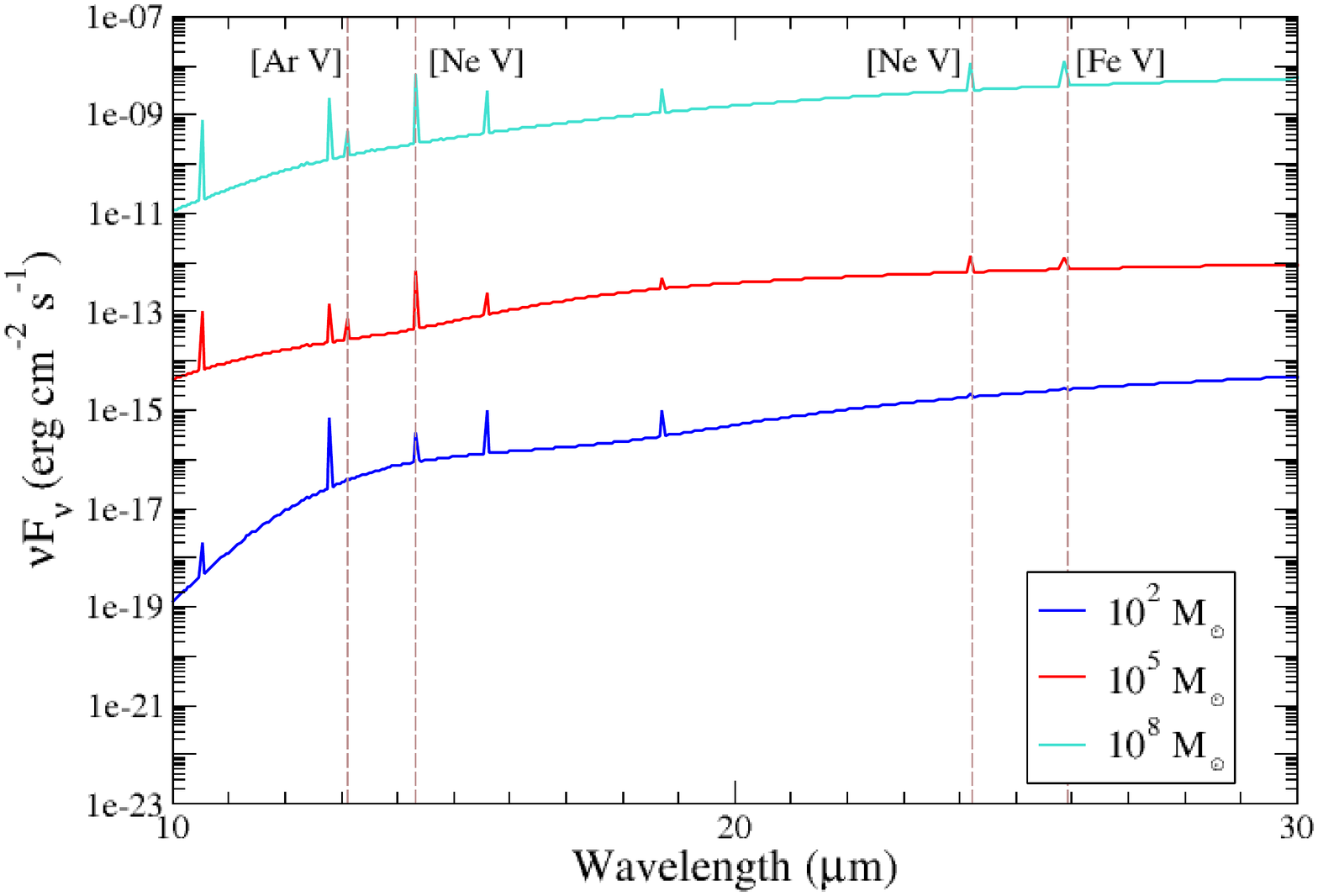}
\caption{Simulated spectra between $1-30$ $\mu$m for black holes of mass $10^2$M$_{\odot}$, $10^5$M$_{\odot}$, and $10^8$M$_{\odot}$ for a standard gas density of $n_{H}=300~\rm{cm^{-3}}$ and an ionization parameter of $\log$U=-1.0. Key emission lines are labeled in the figure.}
\label{figure_label3}
\end{figure}

\begin{figure}

\includegraphics[width=0.45\textwidth,height=5.5cm]{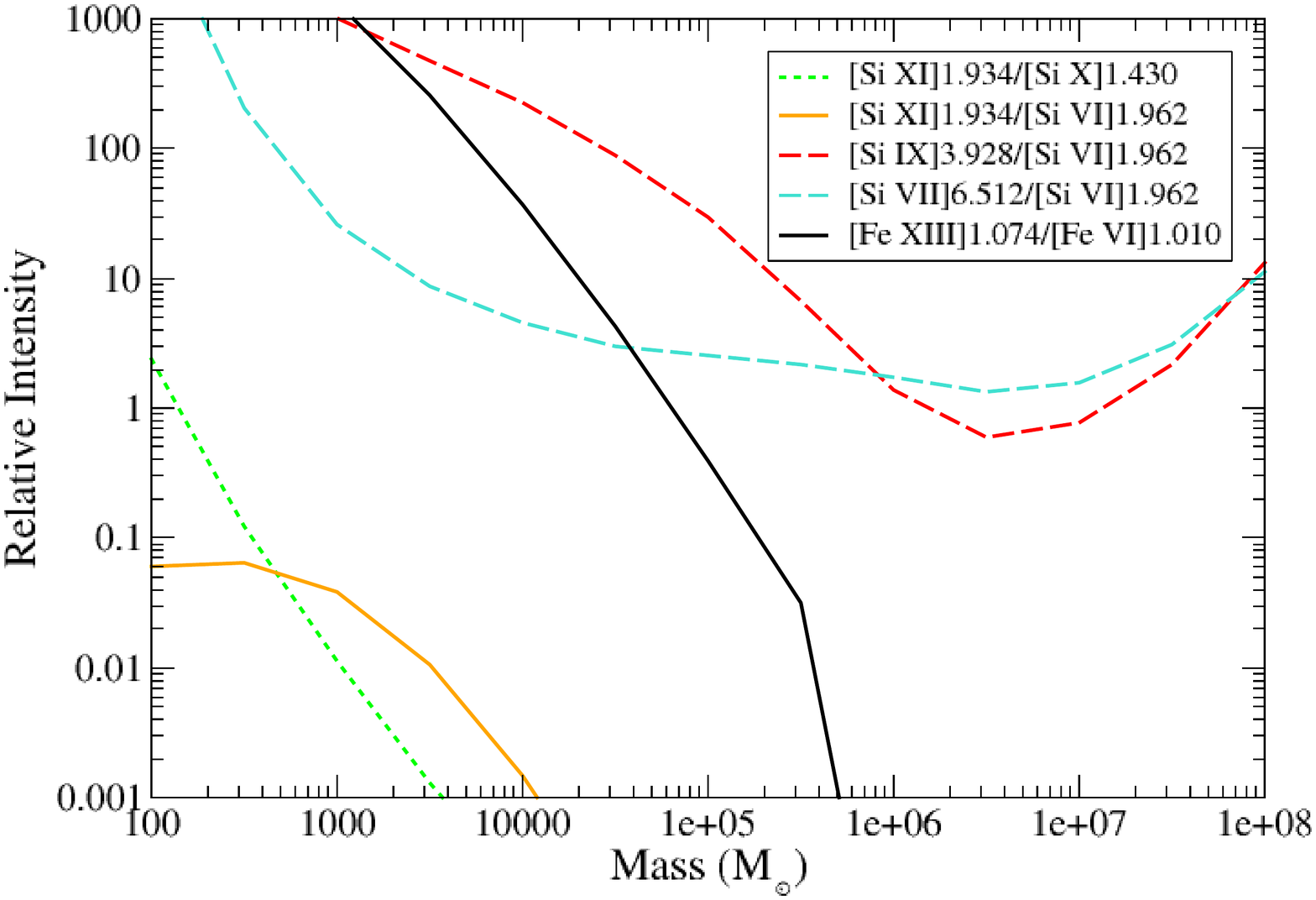}
\includegraphics[width=0.45\textwidth,height=5.5cm]{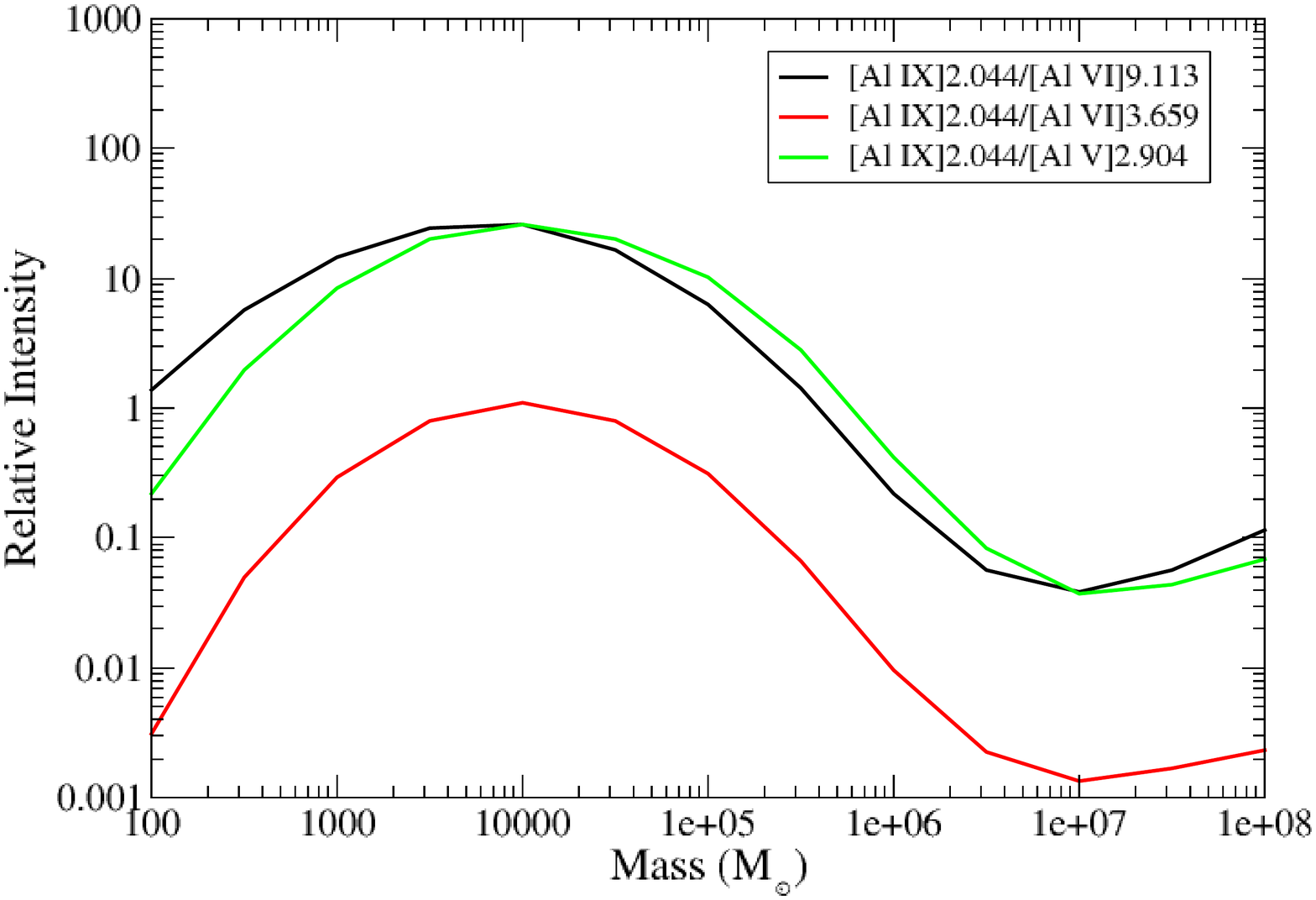}
\includegraphics[width=0.45\textwidth,height=5.5cm]{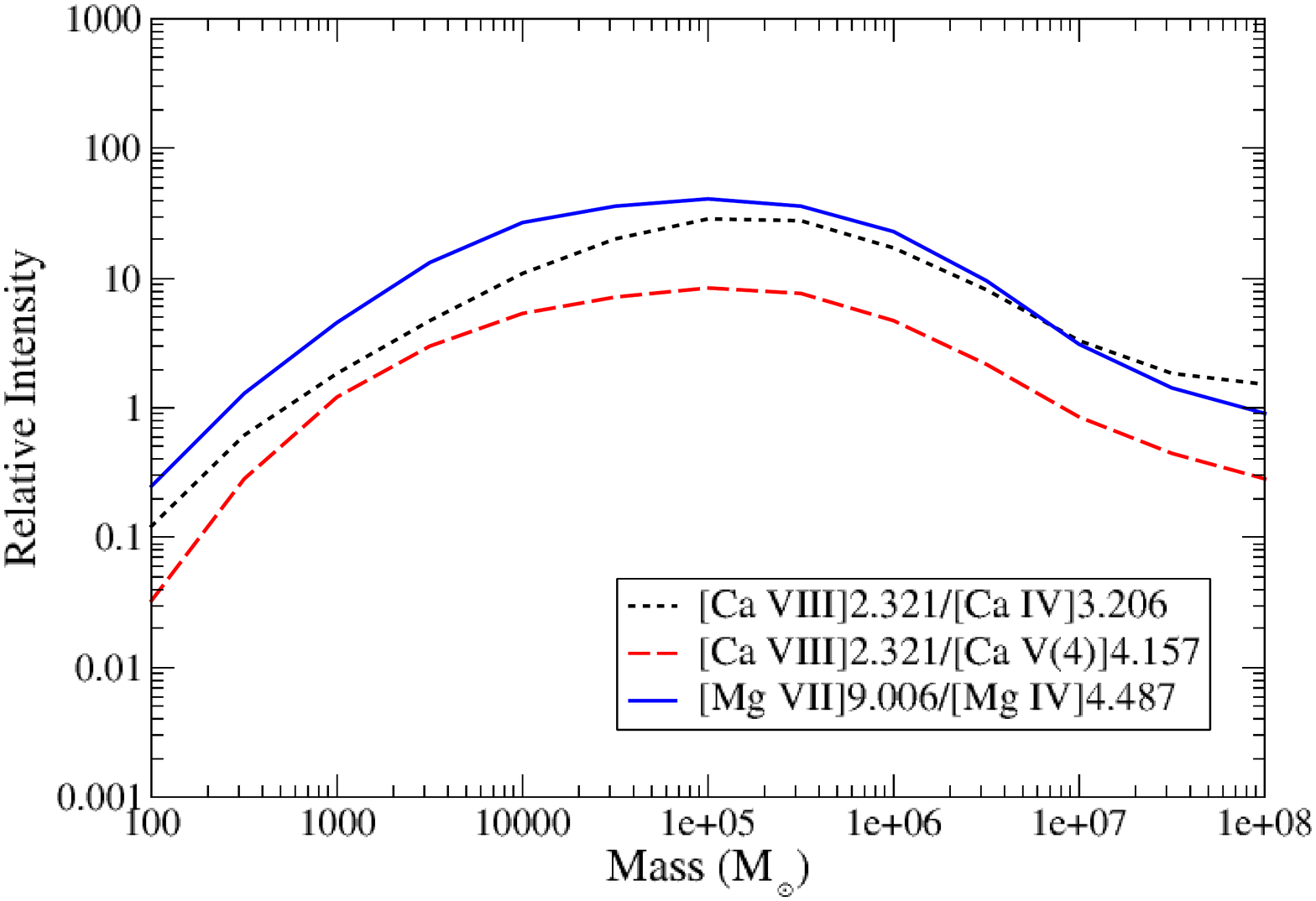}

\caption{Infrared CL ratios as a function of black hole mass for our $n_{H}=300~\rm{cm^{-3}}$ and $\log$U= $-1.0$ models (top) and logU = $-2.0$ (middle and bottom). Note that these diagrams are not meant to be compared to observations.}
\label{figure_label4}
\end{figure}

The predicted infrared emission line spectrum in our simulations is strongly dependent on the shape of the ionizing radiation field incident on the gas and hence the black hole mass. In Figure ~\ref{figure_label3}, we show the transmitted continuum in the $1-30$ $\mu$m range for the $10^2$ M$_{\odot}$ and $10^8$ M$_{\odot}$ AGN models for a standard gas density of $n_{H}=300$ cm$^{-3}$ and an ionization parameter of $\log$U = $-1.0$. As can be seen, the strength of the emission lines in this wavelength range changes significantly with black hole mass, with the CLs with the highest ionization potentials being more prominent in the lower black hole mass model (see Table \ref{tab1}). In Figure ~\ref{figure_label4}, we show the mass dependence of a selection of key abundance-insensitive line ratios as a function of black hole mass, again for $n_{H}=300$ cm$^{-3}$ and $\log$U = $-1.0$. As can be seen, for a given ionization parameter and gas density, line ratios involving ions with different ionization potentials vary dramatically with black hole mass in response to changes in the shape of the AGN ionizing continuum with black hole mass. The CLs with the highest ionization potentials ($\gtrapprox$ 300 eV) peak at the lowest black hole masses, and the line ratios can vary by over six orders of magnitude between $10^2$M$_{\odot}$ and $10^6$M$_{\odot}$.  We also find a series of line ratios that peak in the intermediate mass range (10$^4 -$ 10$^5$M$_{\odot}$), demonstrating that the suite of line ratios is sensitive to different mass ranges explored in our models.  

\par

\begin{figure*}[]

\centering

\begin{tabular}{cc}

\includegraphics[width=0.42\textwidth]{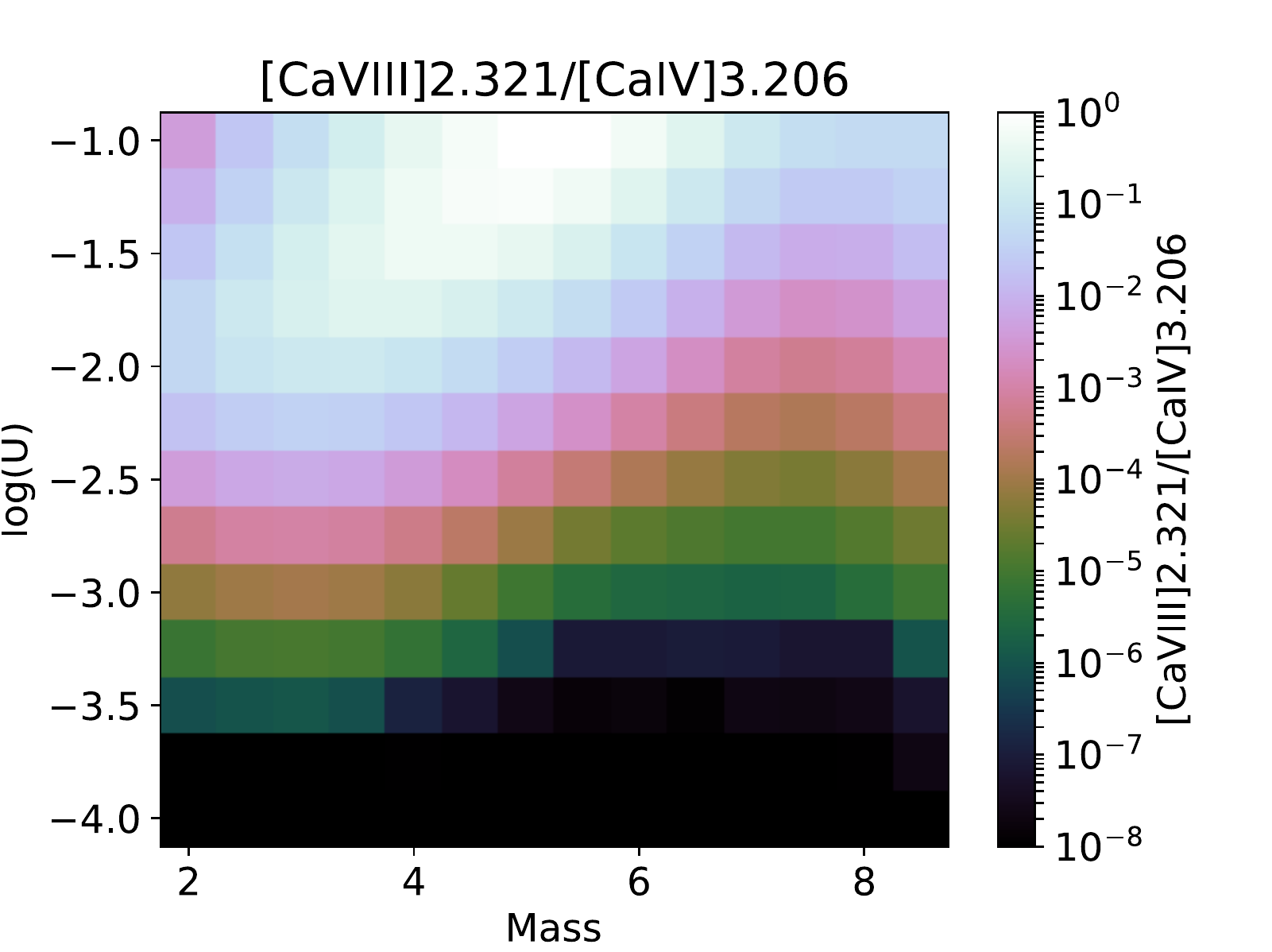} & \includegraphics[width=0.42\textwidth]{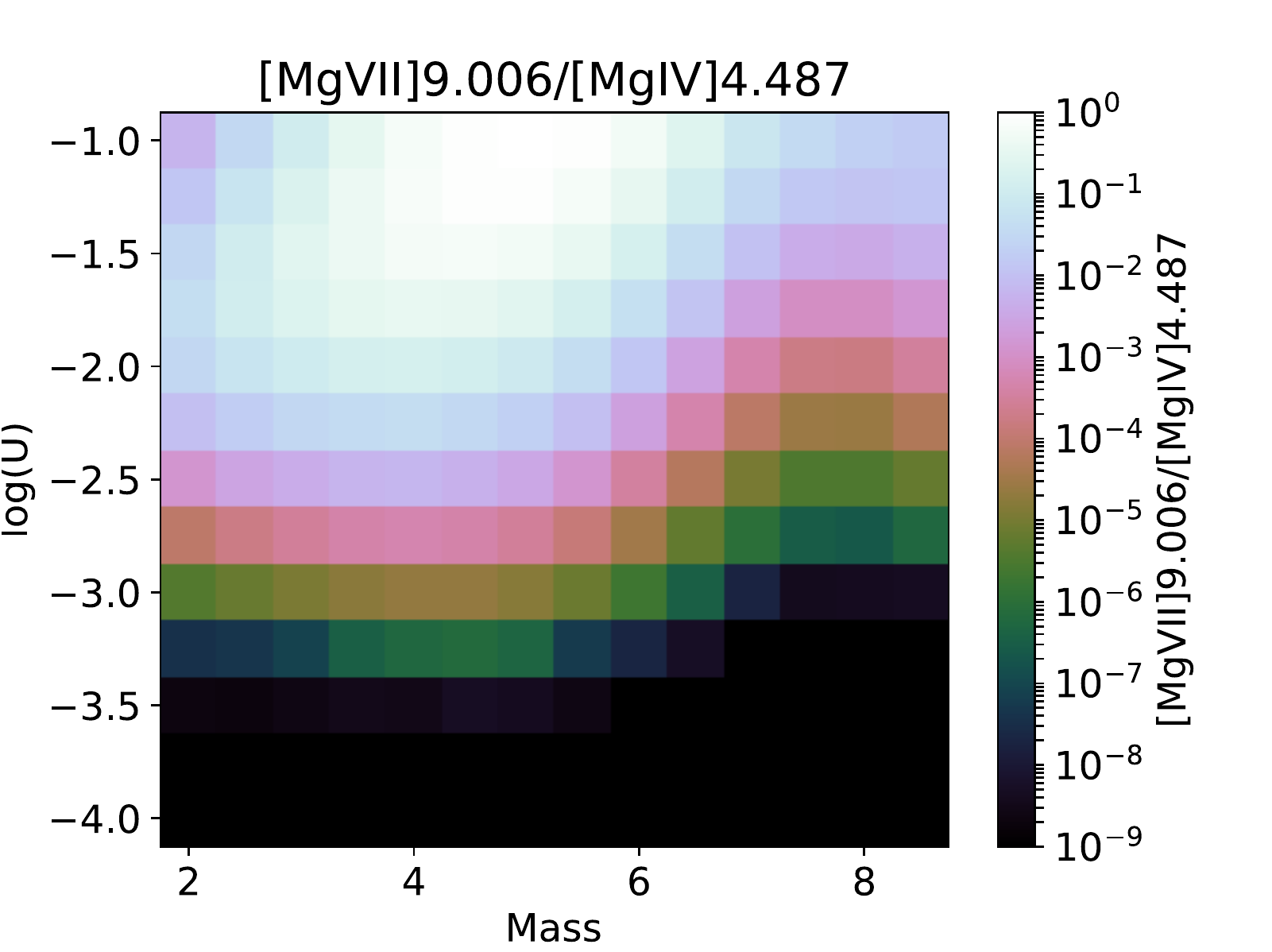} \\

\includegraphics[width=0.42\textwidth]{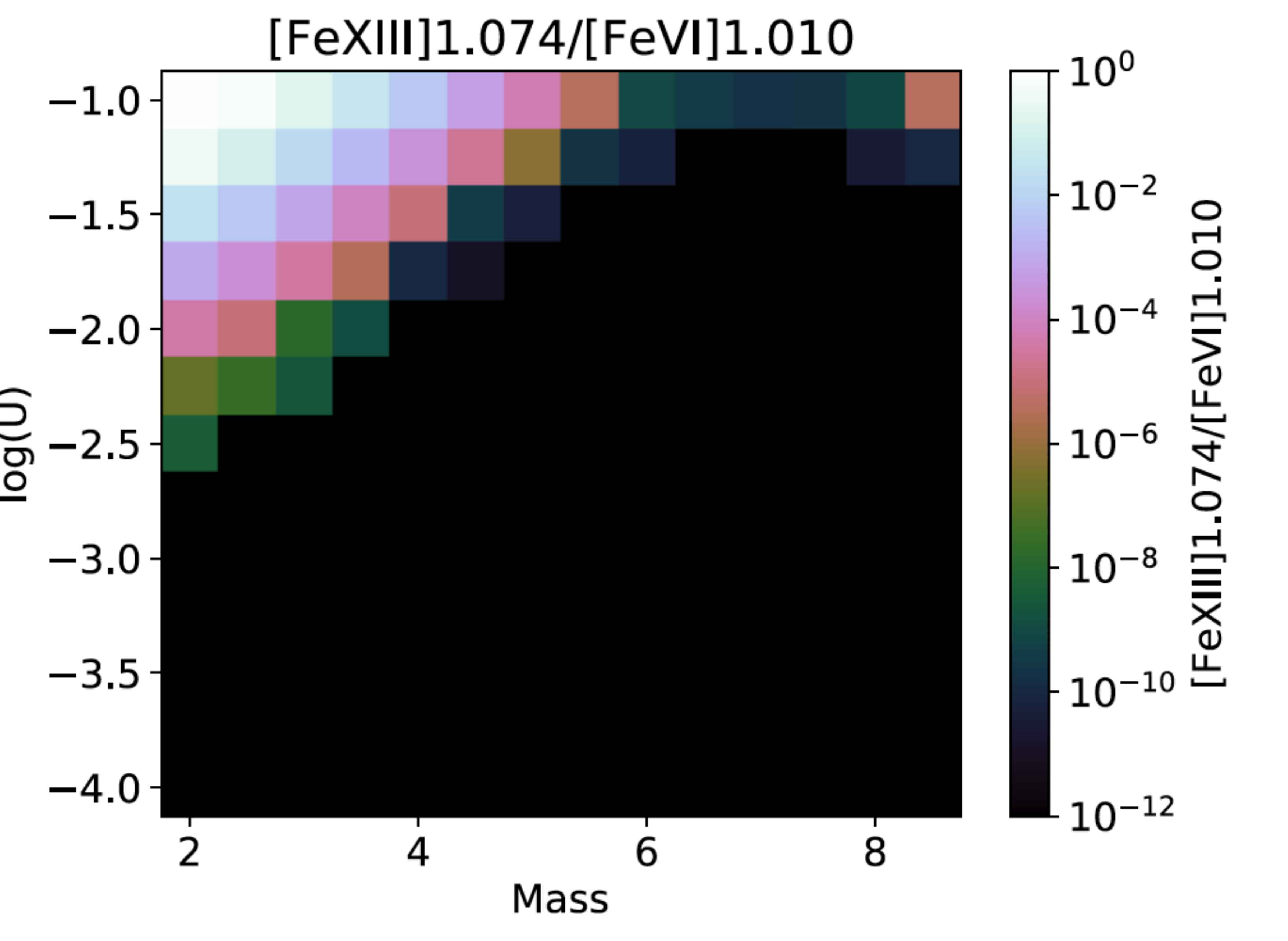} & \includegraphics[width=0.42\textwidth]{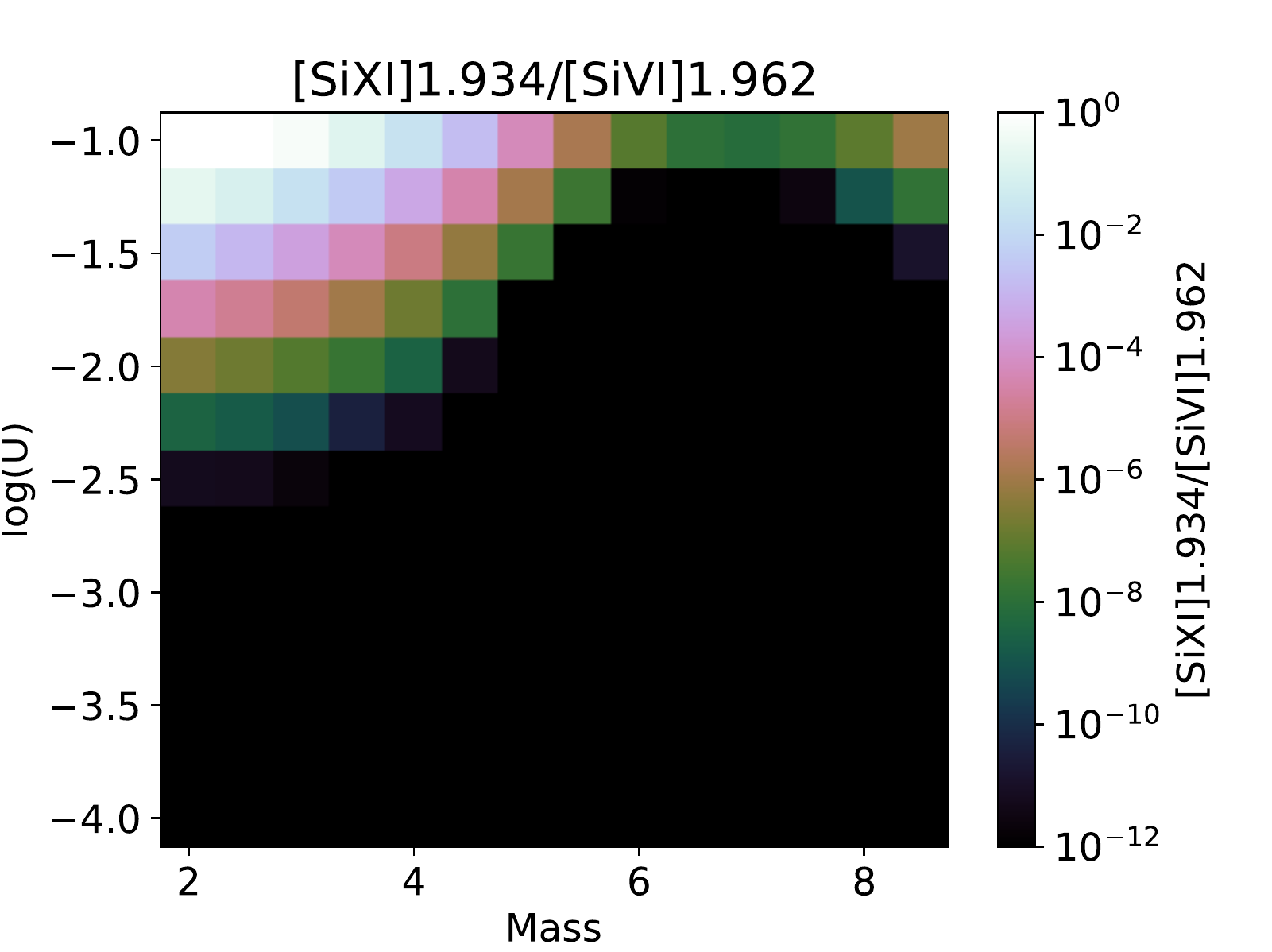} \\

\end{tabular}
\caption{Contour plots showing the dependence on mass and U at $n_{H}=300~\rm{cm^{-3}}$ for various predictive line ratios.  While high ratios generally are indicative of a definitive mass range, lower ratios can correspond to many possibilities. The highest ratio in each plot is normalized to one.}
\label{figure_label5}
\end{figure*}

\begin{figure*}[]

\centering

\begin{tabular}{cc}

\includegraphics[width=0.42\textwidth]{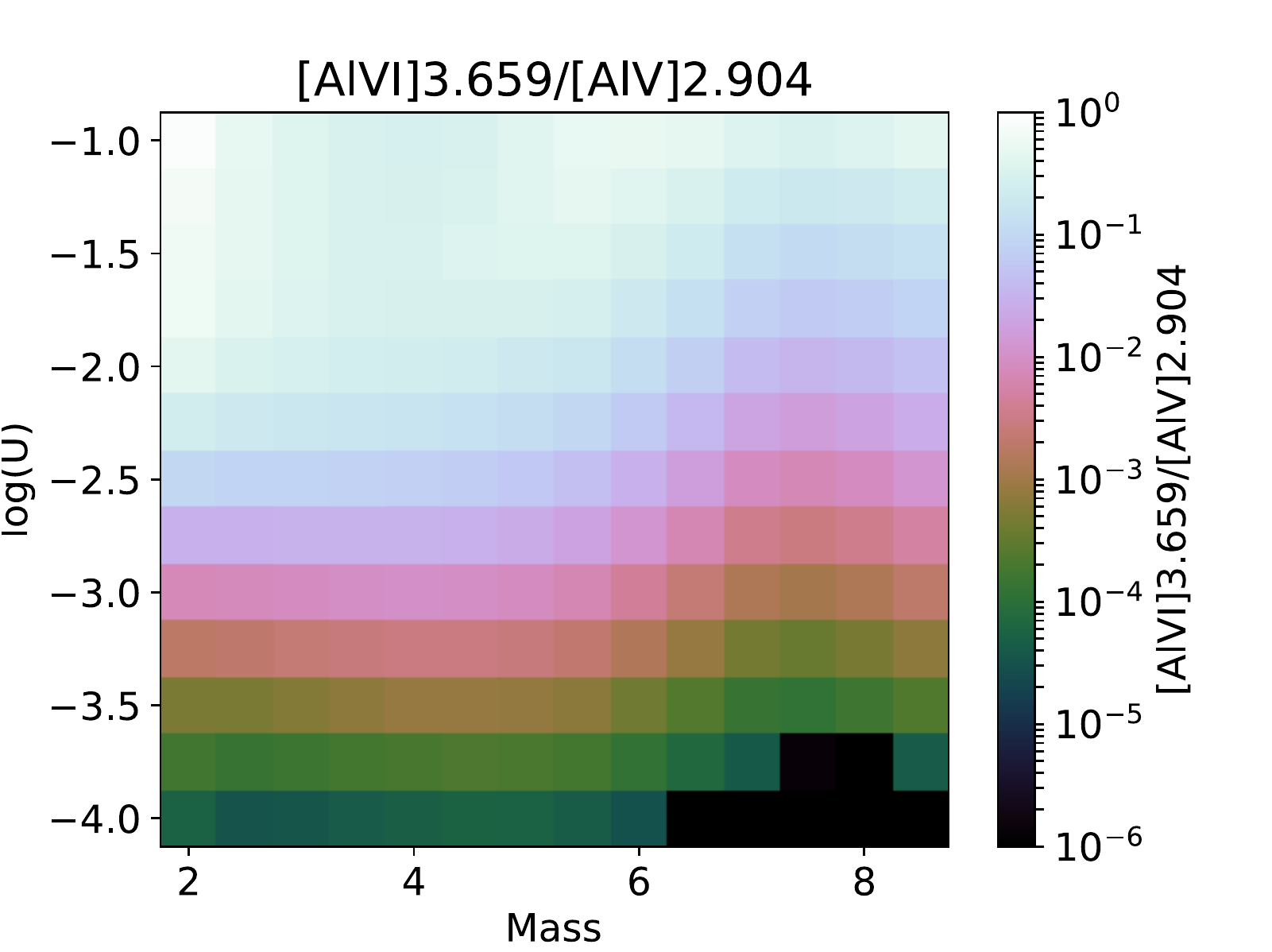} & \includegraphics[width=0.42\textwidth]{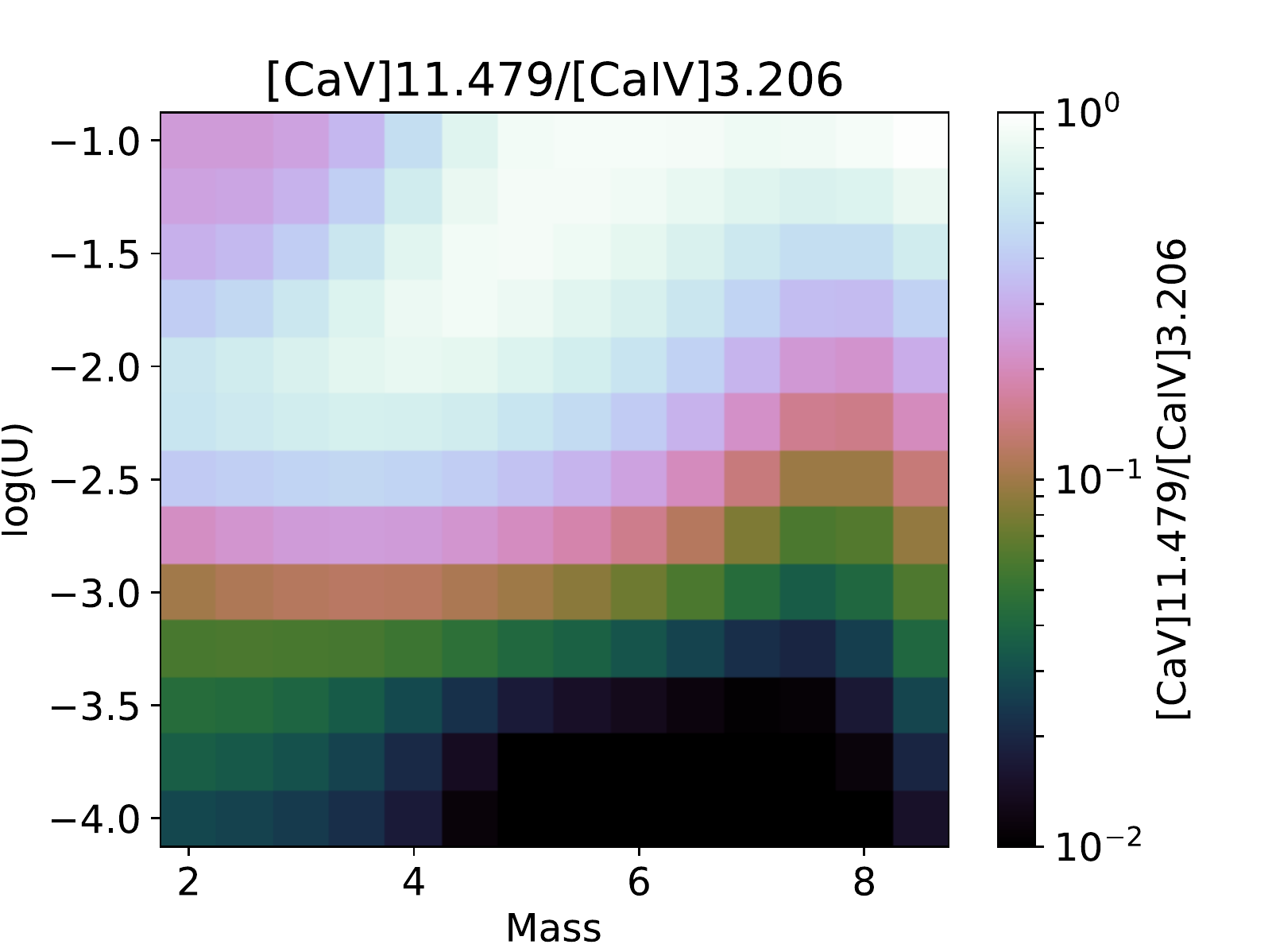} \\

\includegraphics[width=0.42\textwidth]{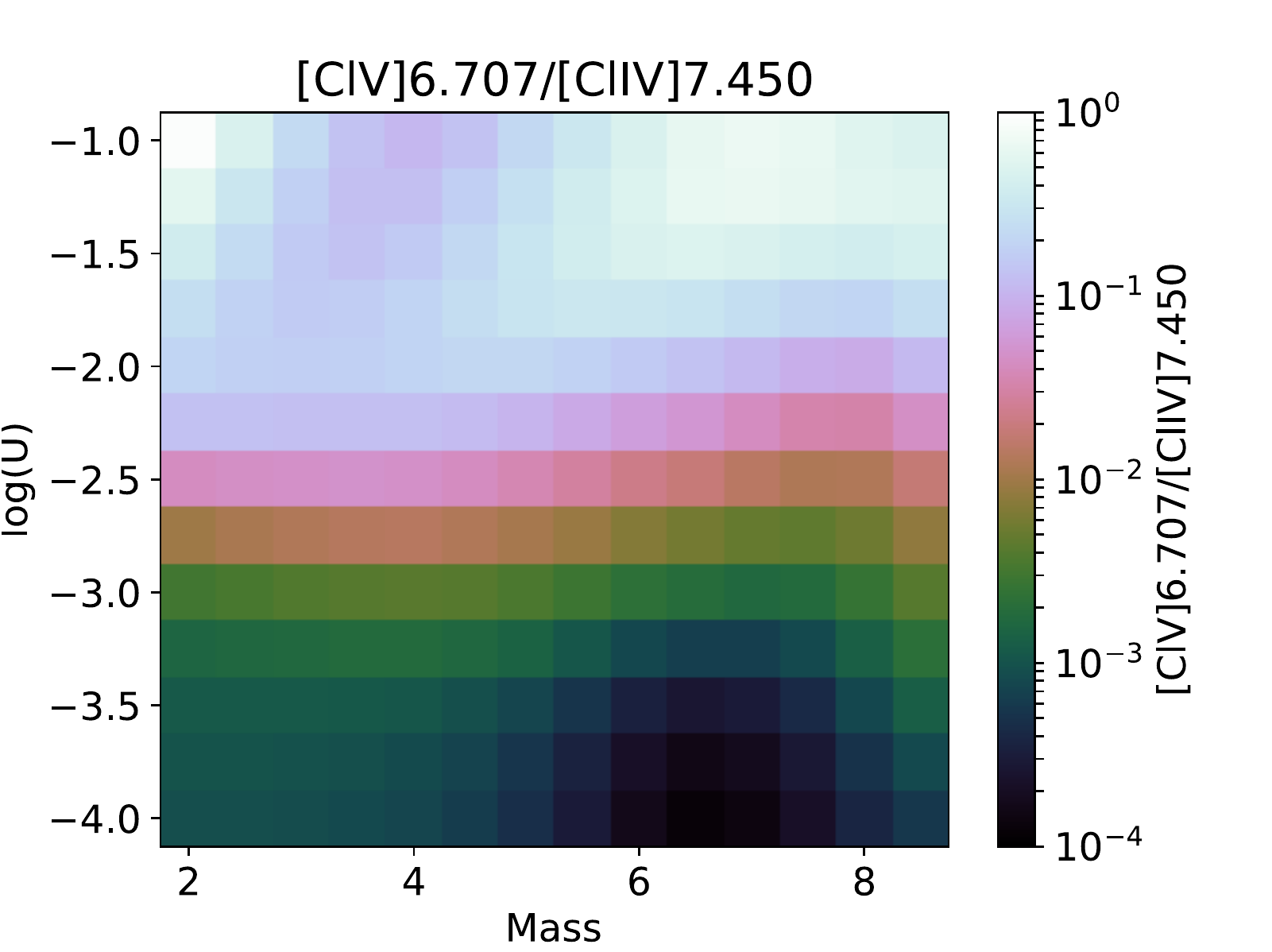} & \includegraphics[width=0.42\textwidth]{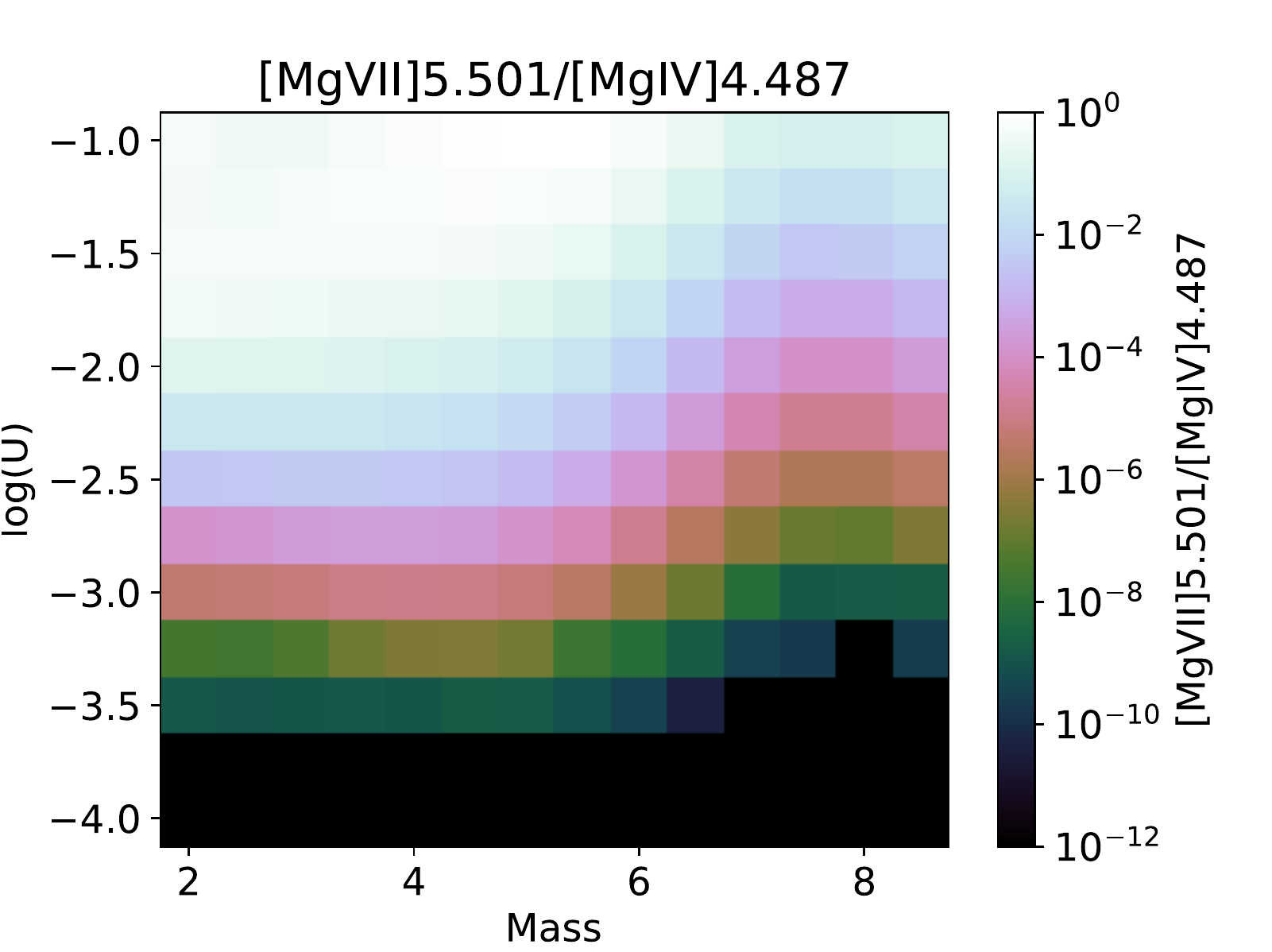} \\

\end{tabular}
\caption{Contour plots showing the dependence on mass and U at $n_{H}=300~\rm{cm^{-3}}$ for a selection of line ratios more sensitive to U than black hole mass. As can be seen, these CL ratios vary more significantly with U than with black hole mass over the range of parameter space explored in our models.  The highest ratios in each plot is normalized to one.}
\label{figure_label6}
\end{figure*}

The CL ratios vary with both the shape of the ionizing continuum and the ionization parameter adopted in our models. As the ionization parameter increases, the ratio of photon flux to gas density increases, which in turn increases the ionization rate while keeping the recombination rate relatively constant for our constant density models. The combination of the hardness of the radiation field, which depends on the black hole mass, and the ionization stage of a particular element determines the CL ratio for a given gas density.  In Figure ~\ref{figure_label5}, we show the dependence of a selection of key CL ratios with both black hole mass and ionization parameter to illustrate this point. As can be seen, the line ratios are sensitive to both the black hole mass and the ionization parameter since both parameters affect the dominant ionization stage of a given species and the ionization structure of the nebula. However, CLs associated with the highest ionization potentials are most prominent only in models with the smallest black hole masses and high ionization parameters, indicating that extremely high observed ratios will be associated only with low mass black holes. For intermediate ratios, our models demonstrate that there are degeneracies between U and black hole mass. In such cases, a large set of CL observations can be used constrain the black hole mass and the ionization parameter. Based on our models, we identify line ratios that display much larger variations with ionization parameter than with mass as can be seen from Figure ~\ref{figure_label6}.  This ratio stays nearly constant across the lower mass range, but spans a wide range of values in ionization parameter, and so could be a potential diagnostic for determining ionization parameter in an AGN.

\section{Discussion}

Based on our photoionization models, we have demonstrated that the infrared CLs are potentially a powerful probe of the black hole mass in AGNs. In particular, emission lines associated with ions with the highest ionization potentials, such as SiXI, SiX, and FeXIII are predicted to be far stronger in the smallest black hole mass models. Our models suggest that the following line ratios will be higher for AGNs powered by IMBHs with masses $<$ $10^6$ M$_{\odot}$:
\begin{itemize}
\item{[SiXI]1.934$\mu$m/[SiX]1.430$\mu$m}
\item{[SiXI]1.934$\mu$m/[SiVI]1.962$\mu$m} 
\item{[SiIX]3.928$\mu$m/[SiVI] 1.962$\mu$m}
\item{[SiVII]6.512$\mu$m/[SiVI]1.962$\mu$m}
\item{[FeXIII] 1.074$\mu$m/[FeVI]1.010$\mu$m}
\end{itemize}
which are most prominent for the lowest mass IMBHs ($<$ $10^4$M$_{\odot}$), and the
\begin{itemize}
\item{[AlIX]2.044$\mu$m/[AlVI]9.113$\mu$m}
\item{[AlIX]2.044$\mu$m/[AlVI]3.659$\mu$m} 
\item{[AlIX]2.044$\mu$m/[AlV]2.904$\mu$m}     
\item{[CaVIII]2.321$\mu$m/[CaIV]3.206$\mu$m} 
\item{[CaVIII]2.321$\mu$m/[CaV]4.157$\mu$m} 
\item{[MgVII]9.006$\mu$m/[MgIV]4.487$\mu$m}
\end{itemize}
which peak for IMBHs in the $10^4$M$_{\odot}-10^6$M$_{\odot}$ range.

These line ratios are predicted to vary by many orders of magnitude with black hole mass and can thus help inform future AGN studies with {\it JWST}.
\par

Based on existing observations of AGNs, the number of CLs detected decreases with increasing ionization potential \citep{rodriguez2006}. Indeed, in the entire sample of 838 nearby powerful AGNs from the {\it Swift/BAT} survey, not a single [SiXI] and [S XI] line is detected in the 102 observed in the near-IR \citep{lamperti2017}. We searched the literature for all CL studies of nearby AGNs and found that the most frequently detected CL is the [SiVI]1.962~$\mu$m line, which is detected in a total of 76 AGNs \citep{lamperti2017, mason2015, muller-sanchez2007, riffel2006, rodriguez2011, mould2012, alonso-herrero2000, rhee2005}. In contrast, we could only find a total of 3 detections of the [SiXI] 1.932~$\mu$m line reported in the literature. Given that the mean mass of the active black holes observed in the literature, measured either through broad lines or the velocity dispersion of the Ca II triplet or the CO band-heads, is $\approx 10^7$M$_{\odot}$ with a very narrow spread in mass, it is possible that this observational fact is caused by a powerful selection effect: we are observing lines that are predicted to be the strongest precisely for the black hole masses probed by our observations. The CLs associated with higher ionization potentials may be weak because we have not yet observed a significant sample of AGNs powered by lower mass black holes. It is possible that these CLs will be enhanced relative to the lower ionization potential CLs in AGNs with lower mass black holes that will be observed with {\it JWST}. 
\par

\begin{figure}
\includegraphics[width=0.45\textwidth]{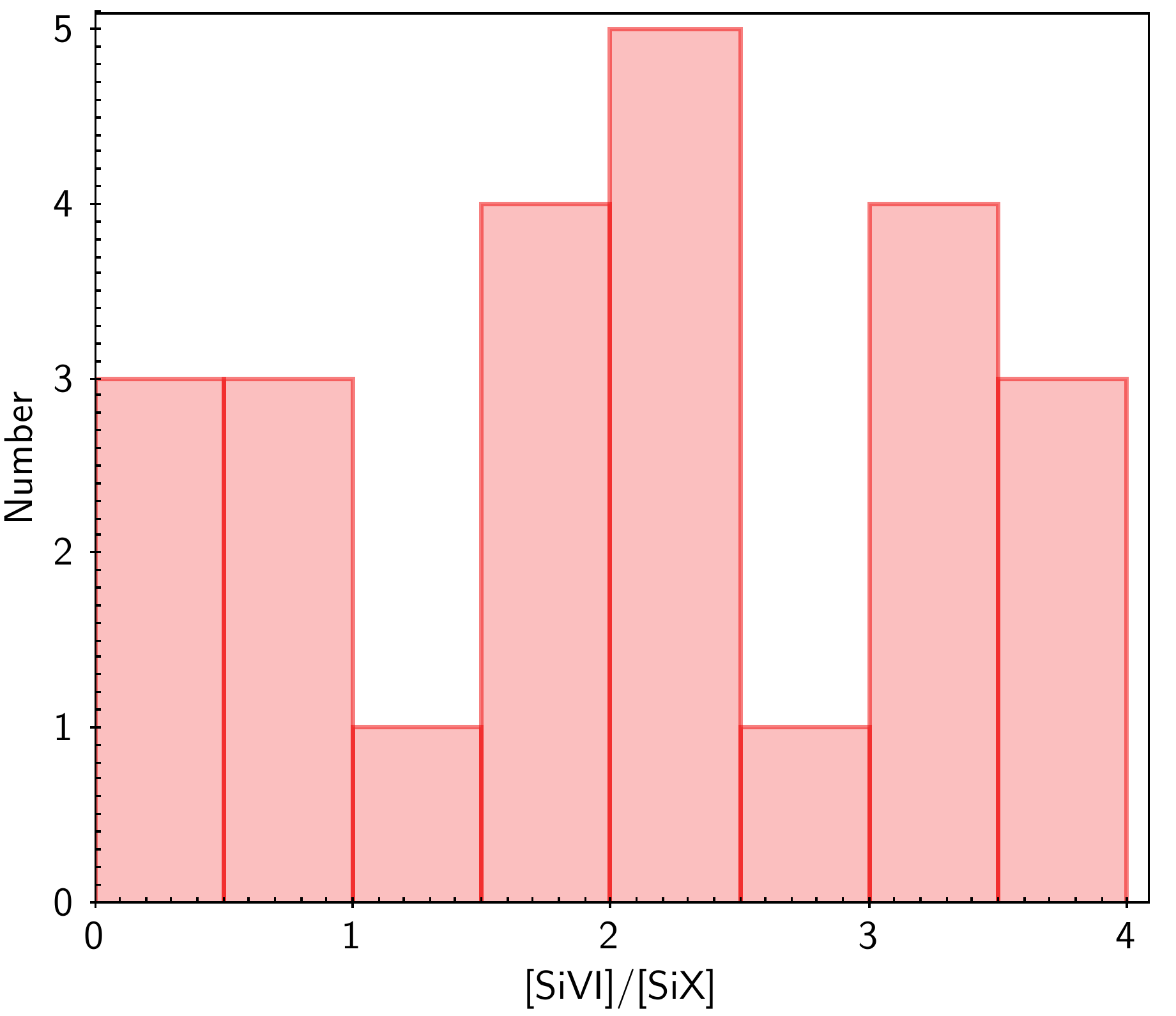}
\caption{Observed [SiVI]1.962/[SiX]1.430 line flux ratios from \citet{lamperti2017, mason2015, muller-sanchez2007, riffel2006, rodriguez2011, mould2012, alonso-herrero2000, rhee2005}}
\label{figure_label7}
\end{figure}

\begin{figure}
\includegraphics[width=0.45\textwidth]{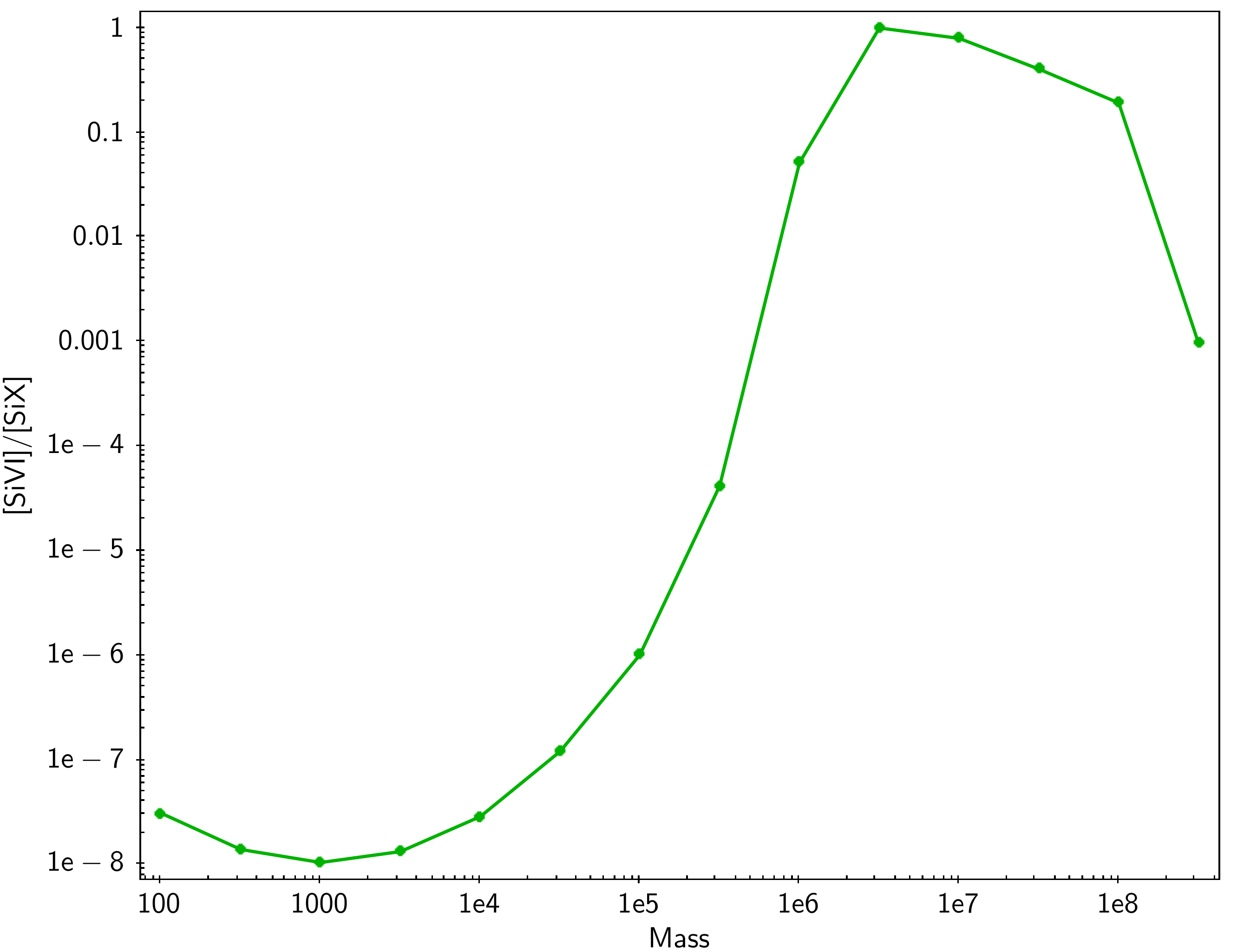}
\caption{Predicted [SiVI]1.962/[SiX]1.430 line flux ratios as a function of black hole mass for a gas density of $n_{H}=300~\rm{cm^{-3}}$ and an ionization parameter of logU= $-2.0$. The line ratio has been normalized to 1.0 at its peak.  }
\label{figure_label8}
\end{figure}

We have shown that several key diagnostic CL line ratios (see Figure ~\ref {figure_label4}) vary by many orders of magnitude over the mass range explored in this work. The observed ratios of detected CLs in the literature thus far do not show this range of variation. In Figure \ref{figure_label7}, we show the distribution of [SiVI]1.962/[SiX]1.430 line flux ratios from the literature. As can be seen the variation in this line ratio is less than an order of magnitude for the relatively small sample of objects observed thus far, suggesting similar black hole masses, ionization parameters and physical properties of the gas. In contrast, this ratio can vary by over seven orders of magnitude over the mass range explored in our calculations as can be seen from Figure ~\ref{figure_label8}. Moreover, the ratio is expected to peak precisely at the black hole masses probed by current observations, suggesting that the predominance of the [SiVI]1.962 CL relative to the higher ionization potential lines may be a selection effect caused by the range of black hole masses thus far explored.  

\subsection{Additional Considerations and Caveats}
Our goal in this paper is to explore the first order effects of varying black hole mass on the coronal line spectrum of gas ionized by an AGN.  We emphasize that we have not attempted to present a model finely tuned to match the properties of any one particular AGN.  The strength and shape of the continuum in the 100-400 eV range is important in replicating observed CL line ratios. The generic AGN model adopted in this work predicts the {\it relative} behavior of the CL ratios as a function of BH mass, and demonstrates which line ratios are the most sensitive to BH mass. The absolute line ratios are not meant to replicate observed line ratios in any particular AGN.  In particular, the AGN continuum model we have adopted does not produce sufficient photons in the EUV or soft X-rays to generate the observed CL strengths seen in nearby AGNs. This has been also found in several other studies which show that photoionization models with a similar AGN continuum as that adopted in this work fail to produce the observed strengths of the HeII edge in quasars \citep{korista97} and the CL spectrum in Circinus \citep{oliva1994}, unless a harder radiation field is introduced. Moreover, the observed emission line spectrum will likely originate from multiple clouds with differing physical parameters. Indeed, the variation in line width with ionization potential and the blueshifts observed in many CLs \citep{grandi1978,marconi1996,mazzalay2007,rodriguez2006,rodriguez2011,muller-sanchez2011} suggest stratification in the photoionized region producing the CLs and the presence of outflows. Simple photoionization models of a single cloud will therefore not likely replicate observed line ratios.
\par

We note that this first exploratory study of the CL emission line spectrum as a function of black hole mass was carried out assuming an accretion rate of $0.1 \dot m_{Edd}$.  Note that the SED from the accretion disk is a function of the accretion rate adopted.  While changes in the accretion rate will affect the predicted CL spectrum, a $10^8$ M$_{\odot}$ black hole would need to accrete at $10^5 \dot m_{Edd}$ to produce an accretion disk of the same temperature as a $100$ M$_{\odot}$ black hole radiating at $0.1 \dot m_{Edd}$. Therefore, the line ratios that uniquely identify low mass black holes as seen in Figure \ref{figure_label4} are likely to still have diagnostic potential.
\par
The strengths of the emission lines and their ratios can also be affected by other physical processes that we have not included in this preliminary exploration. For example, we have not included the effects of shocks in our calculations, which can result in collisional ionization as well as photoionization effects.  Outflows from the AGN or winds driven by star formation in the surrounding gas can produce shock fronts that ionize the gas and affect the observed emission line spectrum \citep{allen2008,kewley2013}. The presence of blue-shifted line profiles in many CLs is indeed consistent with the presence of outflows \citep[e.g.,][]{marconi1996,mazzalay2007,rodriguez2006,rodriguez2011,muller-sanchez2011}. For fast shocks, the ionizing radiation generated by the cooling of the hot gas behind the shock front can generate ionizing radiation that can even extend into the soft X-rays, causing significant photoionization effects for the coronal line spectrum \citep{allen2008}. Several studies suggest, however, that the main driver of the CL emission is photoionization by the AGN continuum \citep[e.g.,][]{oliva1994,korista97,marconi1996}, and there is some question on whether shocks can produce the required energetics to produce the observed CL line luminosities \citep{wilson1997}, although  some contribution from shocks cannot be ruled out \citep{rodriguez2006,geballe2009}. Note that the strength and shape of the ionizing radiation field produced by shocks is a strong function of the shock velocity. A significant contribution from shocks would be accompanied by high shock velocities which would result in other distinguishing features in the emission line spectrum \citep{allen2008,kewley2013}. 

\par

The physics of the dust can also affect the observed CL spectrum. In addition to being a source of heating and depletion of refractory elements, grains will absorb the ionizing radiation, thereby affecting the ionization structure of the nebula. We considered models that did not include the presence of grains, as well as models that varied the grain abundance with distance from the ionizing source in order to simulate the effect of grain destruction due to sublimation from the hard radiation field. These variations only effected the observed CL ratios by less than an order of magnitude, significantly less than the effect of black hole mass.
\par

In addition, our models have assumed a simple geometrically thin optically thick disk. This assumes that the accretion is radiatively efficient, and viscous heating is balanced by radiative cooling. In some instances, the accretion can be advection-dominated and radiatively inefficient \citep{yuan2014b}, producing a significantly different SED in the range of the ionization potentials of the coronal lines studied in this work. We also note that we do not take into account the effect of black hole spin, which would affect the inner radius of the accretion disk, and hence its temperature, and therefore the shape of the emergent ionizing SED from the disk. Our model also does not take into account radiative transfer through the atmosphere of the disk. 
\par
Finally, we should note that the AGN continuum we have adopted is based on observations of the reprocessed radiation. There is significant uncertainty in the actual ionizing radiation field incident on the gas. We have added an ad hoc soft excess component in our calculations, but since the origin of this component and its dependence on mass, Eddington ratio, or AGN luminosity is unknown, we have not explored how variations in this component would impact the observed line ratios. Indeed, because this component contributes significantly to the ionization of the coronal lines, this could prove to be a significant source of uncertainty. While parameters other than black hole mass will have some effect on the observed CL ratios, we have demonstrated that variations in the black hole mass are predicted to cause variations by many orders of magnitude in key CL ratios which we have identified in Figures \ref{figure_label4} and \ref{figure_label5}.  The fact that the observed CL ratios of the current suite of observations which probe a very narrow range in black hole mass show little variation (see Figure \ref{figure_label7}) may suggest that the physical conditions of the gas are not likely to vary significantly in AGNs.



\section{Conclusions}
In this work, we have modeled the infrared emission line spectrum that is produced by gas photoionized by an AGN radiation field and explored for the first time the dependence of the CL spectrum on the black hole mass over the range of $10^2$M$_{\odot}-10^8$M$_{\odot}$. Our photoionization models assume purely an AGN ionizing continuum, with a standard geometrically thin, optically thick accretion disk with variable black hole mass, a power law component, and a soft excess.  To explore mass dependence, in these models we have assumed a single accretion rate of 0.1 $\dot{m}_{Edd}$.  We assumed a constant density, 1D spherical model with a closed geometry.  We have not included the effects of shocks. Our main results can be summarized as follows:

\begin{enumerate}
\item{We have demonstrated that the infrared CLs are potentially a powerful probe of the black hole mass in AGNs. In particular, emission line ratios involving ions with the highest ionization potentials with respect to those with lower ionization potentials, such as [SiXI]/[SiVI], [SiIX]/[SiVI], and [FeXIII]/[FeVI], vary by as much as seven orders of magnitude over the mass range explored in our calculations, with the highest ratios corresponding to the lowest mass black holes.}\\

\item{Our models suggest that key line ratios for AGNs powered by IMBHs with masses $<$ $10^6$M$_{\odot}$ will be the:
\begin{itemize}
\item{[SiXI]1.934$\mu$m/[SiX]1.430$\mu$m}
\item{[SiXI]1.934$\mu$m/[SiVI]1.962$\mu$m} 
\item{[SiIX]3.928$\mu$m/[SiVI] 1.962$\mu$m}
\item{[SiVII]6.512$\mu$m/[SiVI]1.962$\mu$m}
\item{[FeXIII] 1.074$\mu$m/[FeVI]1.010$\mu$m}
\end{itemize}
which are most prominent for the lowest mass IMBHs ($<$ $10^4$M$_{\odot}$), and the
\begin{itemize}
\item{[AlIX]2.044$\mu$m/[AlVI]9.113$\mu$m}
\item{[AlIX]2.044$\mu$m/[AlVI]3.659$\mu$m} 
\item{[AlIX]2.044$\mu$m/[AlV]2.904$\mu$m}     
\item{[CaVIII]2.321$\mu$m/[CaIV]3.206$\mu$m} 
\item{[CaVIII]2.321$\mu$m/[CaV]4.157$\mu$m} 
\item{[MgVII]9.006$\mu$m/[MgIV]4.487$\mu$m}
\end{itemize}
which peak for IMBHs in the $10^4$M$_{\odot}-10^6$M$_{\odot}$ range.} \\

\item{While variations in the physical parameters of the gas can also affect the CL spectrum, we demonstrate that the effect of black hole mass is likely to be the most dramatic effect over the mass range explored in our models. We note that this first exploratory study of the CL emission line spectrum as a function of black hole mass was carried out assuming an accretion rate of $0.1 \dot m_{Edd}$.  Note that the SED from the accretion disk is a function of the accretion rate adopted.  While changes in the accretion rate will affect the predicted CL spectrum, a $10^8$ M$_{\odot}$ black hole would need to accrete at $10^5 \dot m_{Edd}$ to produce an accretion disk of the same temperature as a $100$ M$_{\odot}$ black hole radiating at $0.1 \dot m_{Edd}$. Therefore, the line ratios that uniquely identify low mass black holes are likely to still have diagnostic potential.}\\
\end{enumerate}

With the unprecedented sensitivity of {\it JWST}, a large suite of CLs will be detected for the first time in numerous galaxies. In the SDSS survey, there are close to half a million dwarf galaxies (M $<$ $10^9$M$_{\odot}$) in which black holes with masses less than $\approx~10^6$M$_{\odot}$ may reside. The CL spectrum of these galaxies will provide important insight into the existence and properties of IMBHs in the local universe.

\section{Acknowledgements}

 J. M. C. gratefully acknowledges support from an NSF GRFP and  a Mason 4-VA Innovation grant. We also gratefully acknowledge the use of the software TOPCAT \citep{Taylor2005} and Astropy \citep{astropy2013}.  We acknowledge financial support from FONDECYT 1141218 (C.R.), Basal-CATA PFB-06/2007 (C.R.), the China-CONICYT fund (C.R.), and CONICYT+PAI Convocatoria Nacional subvencion a instalacion en la academia convocatoria $\mathrm{a\tilde{n}o}$ 2017 PAI77170080 (C.R.).




\end{document}